\documentclass[aps,
 pra,
 amsmath,
 amssymb,
 twocolumn,
 reprint,
 noeprint,
 superscriptaddress,
 ]{revtex4-2}
\usepackage{CJK} 
\usepackage{times}
\usepackage{float}
\usepackage[normalem]{ulem} 
\usepackage{graphicx}
\usepackage{bm}
\usepackage{color}
\usepackage{xcolor}
\usepackage{mathrsfs}
\usepackage[colorlinks,linkcolor=blue,citecolor=blue,urlcolor=blue]{hyperref}
\usepackage{array} 
\usepackage{tabularx}
\usepackage{orcidlink}

\begin{document}
\title{Improving adiabatic quantum factorization via chopped random-basis optimization}

\author{Tianlai Yang~\orcidlink{0009-0004-6575-3633}}
\thanks{These authors contributed equally.}
\affiliation{College of Physics, Nanjing University of Aeronautics and Astronautics, Nanjing 211106, China}
\affiliation{State Key Laboratory of Mechanics and Control for Aerospace Structures, Key Laboratory for Intelligent Nano Materials and Devices of Ministry of Education, and Institute for Frontier Science, Nanjing University of Aeronautics and Astronautics, Nanjing 210016, China}

\author{Mo Xiong~\orcidlink{0009-0009-9722-8370}}
\thanks{These authors contributed equally.}
\affiliation{College of Physics, Nanjing University of Aeronautics and Astronautics, Nanjing 211106, China}
\affiliation{Key Laboratory of Aerospace Information Materials and Physics, MIIT, Nanjing 211106, China}

\author{Ming Xue~\orcidlink{0000-0002-6156-7305}}
\email[Corresponding author: ]{mxue@nuaa.edu.cn}
\affiliation{College of Physics, Nanjing University of Aeronautics and Astronautics, Nanjing 211106, China}
\affiliation{Key Laboratory of Aerospace Information Materials and Physics, MIIT, Nanjing 211106, China}

\author{Xinwei Li~\orcidlink{0009-0005-9677-0011}}
\email[Corresponding author: ]{xinwei.li1991@outlook.com}
\affiliation{Beijing Academy of Quantum Information Sciences, Xibeiwang East Road, Beijing 100193, China}

\author{Jinbin Li}
\affiliation{College of Physics, Nanjing University of Aeronautics and Astronautics, Nanjing 211106, China}
\affiliation{Key Laboratory of Aerospace Information Materials and Physics, MIIT, Nanjing 211106, China}

\date{\today}

\begin{abstract} 
Integer factorization remains a significant challenge for classical computers and is fundamental to the security of RSA encryption. Adiabatic quantum algorithms present a promising solution, yet their practical implementation is limited by the short coherence times of current NISQ devices and quantum simulators. In this work, we apply the chopped random-basis (CRAB) optimization technique to enhance adiabatic quantum factorization algorithms. 
We demonstrate the effectiveness of CRAB by applying it to factor the integers ranging from 21 to 2479, achieving significantly improved fidelity of the target state when the evolution time exceeds the quantum speed limit. Notably, this performance improvement shows resilience in the presence of dephasing noise, highlighting CRAB's practical utility in noisy quantum systems. Our findings suggest that CRAB optimization can serve as a powerful tool for advancing adiabatic quantum algorithms, with broader implications for quantum information processing tasks.
\end{abstract}
\maketitle

\section{Introduction}
Quantum computers hold the potential to solve certain computational problems significantly faster than classical computers. For integer factorization problems, no classical algorithms are currently known to furnish polynomial-time solutions. In 1994, Shor introduced a quantum algorithm that exhibits a remarkable reduction in time complexity in comparison to classical factorization algorithms~\cite{shor1994algorithms,shor1999polynomial}. Subsequent experiments have confirmed the effectiveness of Shor's algorithm \cite{lu2007demonstration,lanyon2007experimental,amico2019experimental}. An alternative approach for implementing factorization algorithms is adiabatic quantum computing (AQC)~\cite{albash2018adiabatic,smelyanskiy2004quantum,
smelyanskiy2002dynamicsquantumadiabaticevolution,aharonov2007regev,aharonov2008adiabatic,roland2002quantum,
sarandy2005adiabatic,wei2006modified,somma2012quantum,hen2014period,simon1997on,garnerone2012adiabatic}, which was pioneered by Farhi et al.~\cite{farhi2001,farhi2002quantumadiabaticevolutionalgorithms} and presents a promising alternative to the standard quantum computing framework~\cite{aharonov2008adiabatic,nielsen2010quantum,kempe2006complexity,mizel2007simple,breuckmann2014space,
gosset2015universal,lloyd2016adiabatic,aharonov2009power}. 
AQC encodes problem solutions within the ground state of a problem Hamiltonian, denoted as $\hat{H}_p$. 
The process begins with a system described by an initial Hamiltonian, $\hat{H}_0$, which is in its ground state. Through adiabatic evolution, $\hat{H}_0$ is gradually transformed into $\hat{H}_p$. The adiabatic theorem~\cite{kato1950adiabatic} ensures the system stays in the ground state, thereby yielding the solution. AQC is intrinsically connected to the preparation of ground states in quantum many-body systems~\cite{albash2018adiabatic,gaitan2012ramsey,goto2016bifurcation}, and has achieved remarkable success in quantum annealing~\cite{santoro2006optimization,das2008colloquium,johnson2011quantum,wen19experimental,keever24Towards}.

Evolution time holds significant importance in AQC, and numerous studies have delved into adiabatic techniques to accelerate this process~\cite{farhi2008how,muthukrishnan2016tunneling,brady2017necessary,rezakhani2009quantum,rezakhani2010intrinsic,
perdomo2011study,crosson2014different,hormozi2017nonstoquastic,seki2012quantum,dickson2011does}. For example, integrating various counter-diabatic (CD) terms into the total Hamiltonian can substantially boost the probability of successfully reaching the ground state~\cite{hegade2021digitized}. 
The construction of CD terms typically requires knowledge of the Hamiltonian spectrum, which is often experimentally infeasible~\cite{guery19rmpshortcut}. 
Local CD driving protocols~\cite{morawetz2024efficient,cepaite2023counterdiabatic,dries2017minimizing,claeys2019floquet,zhou2020experimental,hartmann2019rapid,li2024quantum} offer a practical alternative, as they avoid the need for complete spectral information. Nonetheless, achieving fast evolution through CD methods may still require substantial modifications to the Hamiltonian, such as large-amplitude control fields, which may be impractical on current quantum hardware.

Additionally, quantum optimal control (QOC) theory provides a non-adiabatic approach to quantum control. Recent efforts have explored the integration of QOC with prime factorization tasks~\cite{tesoro2024quantum}. 
Conventional QOC algorithms, including gradient ascent pulse engineering~\cite{khaneja2005optimal,xu2012quantum} and the quantum approximate optimization algorithm~\cite{farhi2014QAOA}, typically focus on optimizing discrete time intervals $T_i$. 
In contrast, the chopped random-basis (CRAB) optimization~\cite{doria2011optimal,caneva2011CRAB} enables the continuous modulation of the Hamiltonian using a finite number of parameters, offering a distinct advantage in the realm of quantum control. It also enables the optimization of quantum evolution without the prerequisite of the Hamiltonian spectrum and consistently converges to the optimized solution.
With on-device quantum-classical feedback~\cite{marciniak2021optimal}, CRAB can be adapted to realistic quantum systems with noisy environments by iteratively measuring observables. The CRAB ansatz for QOC has demonstrated its versatility as a robust tool in various applications of quantum technology, including quantum computing, quantum simulation, and quantum sensing~\cite{matthias2022one}. 

In this work, we reformulate the factorization problem by targeting the ground state of a specific Hamiltonian, as discussed in previous work~\cite{hegade2021digitized}, and implement this approach using adiabatic evolution combined with CRAB optimization.
Our optimization strategy is designed to enhance the fidelity of the final target state within a given evolution time. 
We apply CRAB optimization to factor several composite integers--21, 77, 91, 187, 703, and 2479--and demonstrate competitive performance. 
In particular, the fidelities obtained for 21, 91, 2479 are used for direct comparison with Ref.~\cite{hegade2021digitized}.
We further find that the minimum evolution time required for optimal CRAB  performance closely aligns with the quantum speed limit (QSL)~\cite{caneva2011speeding}. 
When the total evolution time exceeds the QSL, CRAB optimization not only achieves rapid convergence but also substantially enhances fidelity, even in the presence of dephasing noise, highlighting its robustness in noisy quantum environments.

This paper is organized as follows. In Sec.~\ref{sec:AQC}, we introduce the physical process of integer factorization problems constructed based on adiabatic quantum evolution.
Sec.~\ref{sec:crab} presents the application of CRAB optimization in the factorization of the numbers 21, 91, and 2479. It also includes additional examples involving 77, 187, and 703, analyzes the relationship between the minimum evolution time required for CRAB optimization and the quantum speed limit, and demonstrates the robustness of CRAB in the presence of dephasing noise.
Finally, Sec.~\ref{sec:conclusion} concludes with a discussion.

\section{Integer factorization based on adiabatic quantum computing}\label{sec:AQC}
We consider the factorization of an integer $\omega$ as $\omega = ab$, where $a$ and $b$ are its factors. Assuming, without loss of generality, that $\omega$ is an odd integer, then $a$ and $b$ are both odd. 
Following the approach in Ref.~\cite{hegade2021digitized}, 
we construct a time-dependent Hamiltonian
$\hat{H}(t)$, defined as,
\begin{equation}
   \hat H(t)=[1-s(t)]\hat H_0+s(t)\hat H_p\label{eq:Ht},
\end{equation}  
where the final Hamiltonian $\hat H_p$ encodes the solution to the factorization problem.
Here, $s(t)$ is a scheduling function that evolves from $s(0)=0$ (at which $\hat{H}=\hat{H}_0$) to $s(T) = 1$ (at which $\hat{H}=\hat{H}_p$) over the time interval $t \in [0, T]$. 
The initial Hamiltonian $\hat{H}_0$ can be chosen as
\begin{equation}
 \hat H_0=-g(\hat{\sigma}_x^{(1)}+\hat{\sigma}_x^{(2)} + \cdots + \hat{\sigma}_x^{(n)}), \label{eq:5} 
\end{equation}
which describes $n$ spins interacting with a magnetic field along the $x$-direction with strength $g$. 
Here $\sigma_{x,y,z}^{(i)}$ denotes the Pauli operators for the $i$-th qubit.
The ground state of $\hat H_0$ is given by $|\psi_g(0)\rangle=[(|0\rangle+|1\rangle)/\sqrt{2}]^{\otimes n}$, which can be readily prepared experimentally. 
The state $|\psi_g(0)\rangle$ evolves into the ground state $|\psi(T)\rangle$ of $\hat{H}_p$, 
according to the adiabatic theorem for a sufficiently long time $T$, governed by the Schr\"{o}dinger equation $i\partial_t|\psi(t)\rangle= \hat H(t)|\psi(t)\rangle$. 
The optimization process focuses on finite evolution time $T$ by adjusting the parameter $s(t)$. 
There are two approaches to constructing the problem Hamiltonian $\hat H_p$: {\it direct optimization} method and the {\it binary-multiplication-table} method combined with classical preprocessing.

For the direct optimization method, we define the cost function $f(a,b)$ as:
\begin{equation}
f(a,b) = (\omega - ab)^2,\label{eq:fab=0}
\end{equation}
which minimizes to $f(a,b) = 0$ when $ab=\omega$, thus solving the factorization problem. Our goal is to identify the values of $a$ and $b$ that minimize the cost function $f(a,b)$ by constructing the Hamiltonian based on Eq.~\eqref{eq:fab=0}. 
We represent the integers $a$ and $b$ in binary form:
\begin{align}
a&=a_0+2a_1+2^2a_2+\cdots+2^{n_a}a_{n_a}, \nonumber\\
b&=b_0+2b_1+2^2b_2+\cdots+2^{n_b}b_{n_b}.   \nonumber
\end{align}
Since both $a$ and $b$ are odd, $a_0$ and $b_0$ are set to 1, with $a_l$ (for $l\in[1,2,\cdots,n_a]$) and $b_m$ (for $m\in[1,2,\cdots,n_b]$) being either 0 or 1. The number of qubits required to represent $a$ and $b$ is given by:
$n_a=m(\lfloor \sqrt\omega \rfloor_o)-1$,
$n_b=m(\lfloor \frac \omega3 \rfloor)-1$,
where $\lfloor x \rfloor$ (and $\lfloor x \rfloor_o$) denotes the largest (odd) integer less than or equal to $x$, 
and $m(y)$ indicates the smallest number of bits required to represent $y$~\cite{peng08quantum,hegade2021digitized}. 
The problem Hamiltonian $\hat{H}_p$~\cite{andriyash2016boosting,dridi2017prime,jiang2018quantum,peng2019factoring,mengoni2020breaking,
wang2020prime,ding2024effective}, encoding the solution for factorization, is given by:
\begin{equation}
  \hat H_p =\left[\omega \hat{I}-(\hat{I}+\sum^{n_a-1}_{l=1}2^l\hat{a}_l)\!\otimes\!
  (\hat{I}+\sum^{n_b-1}_{m=1}2^m\hat{b}_m)\right]^2, \label{eq:Hp}
\end{equation}
where $\hat{a}_l = (\hat{I}-\hat{\sigma}^{(l)}_z)/{2}$, $\hat{b}_m = \frac{\hat{I} - \hat{\sigma}^{(m)}_z}{2}$,
and $\hat{I}$ is the identity operator. 
The ground state of $\hat H_p$, represented as $|a_1a_2\dots a_{n_a} b_1b_2\dots b_{n_b}\rangle$ in the computational basis, directly corresponds to the binary encoding of the integer factorization problem. The solution is obtained by calculating the nearest integer of $\langle\psi(T)|\hat{a}_l(\hat{b}_m)|\psi(T)\rangle$, which gives the binary representation of $a$ and $b$ by comparing it with 0 and 1. 

\begin{table}
 \caption{Multiplication table for 21 in binary}
  \small
  \centering  
  \begin{ruledtabular}
  \begin{tabular}{cccccc}
     &$2^4$&$2^3$&$2^2$&$2^1$&$2^0$ \\
  $a$&     &     &    1&$a_1$& 1 \\
  $b$&     &     &     &    1& 1 \\
  \colrule
     &     &     &    1&$a_1$& 1 \\
     &     &    1&$a_1$&    1&   \\
 carries&&$c_{2,3}$&$c_{1,2}$& &\\
   \colrule
   $ab=21$& 1 & 0 & 1 & 0 & 1
  \end{tabular}
  \end{ruledtabular}\label{table:multi_21}
\end{table}

However, as $\omega$ increases, the complexity of the problem Hamiltonian grows, making experimental implementation challenging for large $\omega$. To reduce this complexity, one can use a binary multiplication table for classical preprocessing~\cite{schutzhold06adiabatic,schaller2007role}.
This method starts with a binary multiplication table, exemplified in Table~\ref{table:multi_21} for $\omega=21$.
The first two rows represent the bits of the multipliers $a_i$ and $b_i$, followed by four rows showing the intermediate multiplication results. The carries from the $i$-th bit to the $j$-th bit are represented by $c_{i,j}$, and the last row shows the binary representation of $\omega$.
By summing the columns, we get a set of factorization equations and
simplify these equations with binary logical constraints to reduce the qubit requirement. Finally, we construct the problem Hamiltonian by mapping the binary variables to operators on qubits.

\section{Quantum optimization using CRAB} \label{sec:crab}
\subsection{The CRAB algorithm } 
Unlike conventional orthogonal bases, CRAB employs chopped-random bases to express functions.
The introduction of randomness enables a more comprehensive exploration of the function space, thereby enhancing the optimization results~\cite{caneva2011CRAB, doria2011optimal, muller2022one,pagano24role}. In the context of integer factorization, our objective is to find the optimal time-dependent function $s(t)$ that evolves the system from the initial ground state $|\psi_0\rangle$ of the Hamiltonian $\hat{H}_0$ to the target ground state $|\psi_p\rangle$ of the Hamiltonian $\hat{H}_p$ within a finite time $T$.
We define the time-dependent function $s^{\rm CRAB}(t)$ by multiplying the initial guess $s_0(t)$ with a weighting function $f(t)$:
\begin{equation}
s^{\rm CRAB}(t)=s_0(t)f(t),\label{14}  
\end{equation}
where the weighting function $f(t)$ is expanded using orthogonal bases as follows:
\begin{equation}
 f\left(t\right)=1+{\sum_{k=1}^{N_c}\left[A_k\sin(\omega_{k}t)+B_k\cos(\omega_{k}t)\right]}/{\lambda\left(t\right)}.\label{eq:ft}
\end{equation}
Here, $\lambda(t)$ is a time-dependent function designed to satisfy the boundary condition $\lambda(t)\rightarrow\infty$ as $t\rightarrow 0$ and $t\rightarrow T$.
$N_c$ represents the number of groups of optimizing parameters, with the remaining parameters being $\{\omega_{k}, A_k, B_k\}$.
Although the frequencies $\{\omega_k\}$ can be considered free variables in principle, 
it is often more efficient to fix them and minimize over $\{A_k\}$ and $\{B_k\}$.
This transforms the problem from functional minimization to multi-variable minimization~\cite{doria2011optimal,caneva2011CRAB}.
The frequencies are chosen as $$\omega_{k}=2\pi k(1+r_{k})/T, \quad k=1, 2 \cdots N_c,$$ 
where $r_k$ is a random number uniformly distributed within the interval $[-0.5, 0.5]$.
The random perturbations introduced by $r_k$ broadens the search space without increasing the number of optimization parameters. This helps CRAB avoid local minima. In cases where there is no clear physical intuition for selecting the function basis, the CRAB algorithm excels at identifying optimal driving fields~\cite{caneva2011CRAB}.

\subsection{CRAB optimization for factorization of 21}\label{subsec:factor21}
\begin{figure}
  \centering
  \includegraphics[width=0.97\columnwidth]{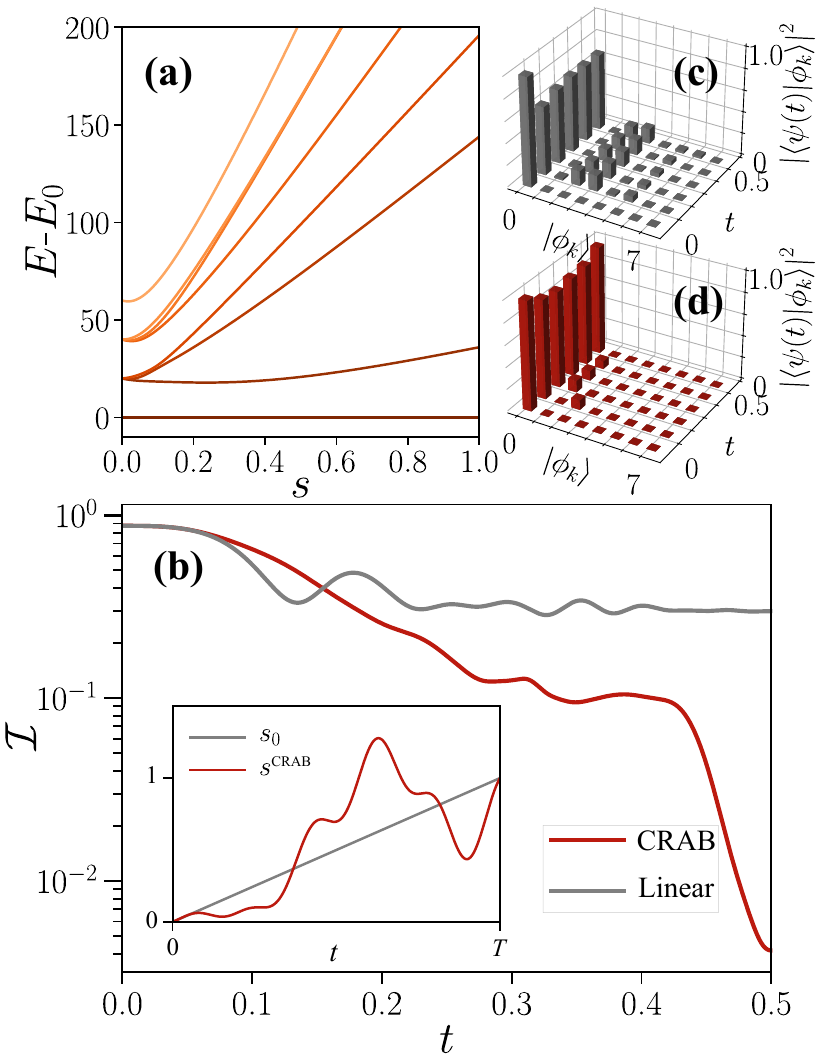}
  \caption{The factorization of 21 before and after CRAB optimization: (a) Spectrum of the total Hamiltonian. The separation of the energy eigenvalues from the lowest energy level $E_0$ is plotted as a function of $s$. (b) The curve of the infidelity $\mathcal{I}$ over time $t$ under the total evolution time $T=0.5$ before (grey) and after optimization (red). Inset: The corresponding time functions, the grey line $s_0(t)$ represents the initial case, and the red line $s^{\rm CRAB}(t)$ represents the case after the CRAB optimization. (c) The projection of instantaneous states on the eigenstates of instantaneous Hamiltonian over evolution using the initial time function $s_0(t)$, and (d) the optimal time function $s^{\rm CRAB}(t)$. 
}
\label{fig:crab21}
\end{figure}
We construct the problem Hamiltonian using the direct optimization method introduced in Sec.~\ref{sec:AQC}. The problem Hamiltonian $\hat{H}_p$ is given by,
\begin{equation}
  \hat H_p=\left[21\hat{I}-(\hat{I}+2\hat{a}_{1})(\hat{I}+ 2\hat{b}_{1}+ 2^2\hat{b}_{2})\right]^2,\label{eq:Hp21}
\end{equation}
where three qubits are required, with $n_a=m(\lfloor\sqrt{21}\rfloor_o)-1=1$, and $n_b=m(\lfloor\frac{21}{3}\rfloor)-1=2$. 
The initial Hamiltonian is $\hat H_0=-10(\hat\sigma_x^{(1)}+\hat\sigma_x^{(2)}+\hat\sigma_x^{(3)})$, with $g=10$~\cite{peng08quantum}. 
Figure~\ref{fig:crab21}(a) shows the energy spectrum of the Hamiltonian $\hat{H}=(1-s)\hat{H}_0+s\hat{H}_p$ as a function of $s$.
The target state $|\psi_p\rangle=|111\rangle$ is the ground state of $\hat{H}_p$. 
We use the energy of the final state $\left|\psi(T)\right\rangle$ at $t=T$ as the cost function,
\begin{equation}
\mathcal{E}(\{A_k\}, \{B_k\})= \left\langle\psi(T)\right|\hat H_p\left|\psi(T)\right\rangle .\label{eq:cost21}
\end{equation}
The coefficients $\{A_k\}$ and $\{B_k\}$ in $f\left(t\right)$ are optimized to minimize the cost function $\mathcal{E}$. We set $N_c=4$ and choose $\lambda(t)=1/\sin(\pi t/T)$ to ensure the boundary conditions $s^{\rm CRAB}(0)=0$ and $s^{\rm CRAB}(T)=1$. 
Optimization can be performed using gradient-based methods, such as gradient descent~\cite{ruder2016overview}, or gradient-free methods, such as the Nelder-Mead algorithm~\cite{singer2009nelder}. These methods are easily implementable using well-established Python libraries, like \textsc{SciPy}~\cite{virtanen2020scipy} and \textsc{Jax}~\cite{jax2018github}.

Figure~\ref{fig:crab21}(b) compares the performance of the factorization of the number $21$ with and without CRAB optimization. 
Using the initial guess $s_0(t)=t/T$, the infidelity between the final state $|\psi(T)\rangle$ and the target state $|\psi_p\rangle$, $\mathcal{I}=1-\left|\langle\psi_p|\psi(T)\rangle\right|^2$, reaches approximately $0.3$ at $T=0.5$,  as shown by the grey line. 
In contrast, applying the CRAB algorithm leads to a significant improvement, 
reducing the infidelity to about $\mathcal{I}(T)\sim \mathcal{O}(10^{-3})$, 
as depicted by the red solid line. 
The inset of Fig.~\ref{fig:crab21}(b) shows the temporal profile of the optimized function $s^{\rm CRAB}(t)$, with the corresponding coefficients listed in Table~\ref{table:coef_21}. The refined characteristics of the function $s^{\rm CRAB}(t)$, in comparison to the initial guess $s_0(t)$, effectively suppress undesired excitations during the adiabatic transformation of the Hamiltonian. 
This is evidenced by the population distribution $P_k$ of the instantaneous eigenstates $|\phi_k(t)\rangle$ throughout the entire temporal evolution, as shown in Fig.~\ref{fig:crab21}(c) and (d). Here, $P_k=|\langle\phi_k(t)|\psi(t)\rangle|^2$, where $|\phi_k(t)\rangle$ is the $k$-th eigenstate of the Hamiltonian $\hat{H}(t)$, and $|\psi(t)\rangle$ is the evolving state.

\begin{table}[!htp]\caption{CRAB optimized coefficients for factorizing 21}\label{table:coef_21}
\begin{tabular}{@{}lllll@{}}
\toprule
$k$       \qquad\qquad & 1      \qquad \qquad & 2      \qquad \qquad &  3      \qquad \qquad  & 4   \\ \hline
$A_k$     \qquad\qquad & -0.116 \qquad \qquad & -1.093 \qquad \qquad &  0.234  \qquad \qquad  & -0.209 \\ 
$r_{k}$ \qquad\qquad & 0.125  \qquad \qquad & -0.348 \qquad \qquad &  0.013  \qquad \qquad  & -0.032 \\ \hline
$B_k$     \qquad\qquad & 0.166  \qquad \qquad & 0.333  \qquad \qquad &  0.477  \qquad \qquad  & 0.340 \\ 
$r_{k}$ \qquad\qquad & -0.181  \qquad \qquad & 0.417  \qquad \qquad &  0.194  \qquad \qquad  & 0.205 \\ \hline
\end{tabular}
\end{table}

\subsection{CRAB optimization for factorization of 2479}
Next, we consider factoring the number 2479 using CRAB optimization. 
We employ the binary multiplication table method introduced in Sec.~\ref{sec:AQC} to obtain the problem Hamiltonian $\hat{H}_p$. 
For the factors $a$ and $b$, we choose bit-lengths $n_a=7$ and $n_b=6$ respectively. 
Both initial and final qubits are set to 1, since the prime factors are odd. 
The multiplication table used for the factorization of 2479 is shown in Table~\ref{table:multi_2479} in Appendix \ref{app:A}. By summing each column in the table, we derive the factorization equations. These equations are then simplified further to reduce the total qubit number required, using classical preprocessing based on binary logical constraints. The final set of equations is as follows:
\begin{eqnarray*}
a_3b_1-b_1 &=& 0,\\
a_3b_2-b_1 &=& 0,\\
a_3+b_2+c_{7,8}-1&=&0,\\
b_1-b_2-2c_{7,8}+1&=&0,\\
a_3-2b_1b_2-b_1+b_2-1&=&0,
\end{eqnarray*}
The total qubit number required for the calculation is 4, determined by the number of variables in the above equations. It is worth mentioning that the required number of qubits scales with the number to be factored as $\mathcal{O}(n\text{log}n)$~\cite{burges2002factoring}.
To obtain the problem Hamiltonian from these simplified equations, we construct the bitwise Hamiltonians by squaring and summing all the equations. Finally, by mapping the variables to qubit operators, the Hamiltonian corresponding to the factorization of 2479 after classical preprocessing is given by:
\begin{eqnarray}
\hat H_p &=& (\hat a_3 \hat b_1-\hat b_1)^2 +(\hat a_3 \hat b_2-\hat b_1)^2 \nonumber\\
 && + (\hat a_3+\hat b_2+\hat c_{7,8}-\hat I)^2 + (\hat b_1-\hat b_2-2\hat c_{7,8}+\hat I)^2\nonumber\\
 && + (\hat a_3-2\hat b_1 \hat b_2-\hat b_1+\hat b_2-\hat I)^2, \label{2479H_p1}
\end{eqnarray}
where $\hat a_3=\frac{\hat I-\hat \sigma_z^{(1)}}{2}$, $\hat b_1=\frac{\hat I-\hat \sigma_z^{(2)}}{2}$, $\hat b_2=\frac{\hat I-\hat \sigma_z^{(3)}}{2}$, 
$\hat c_{7,8}=\frac{\hat I-\hat \sigma_z^{(4)}}{2}$. The ground state of the Hamiltonian $\hat H_p$ corresponds to the solution of factorizing 2479. 

\begin{figure}
  \centering
  \includegraphics[width=0.92\columnwidth]{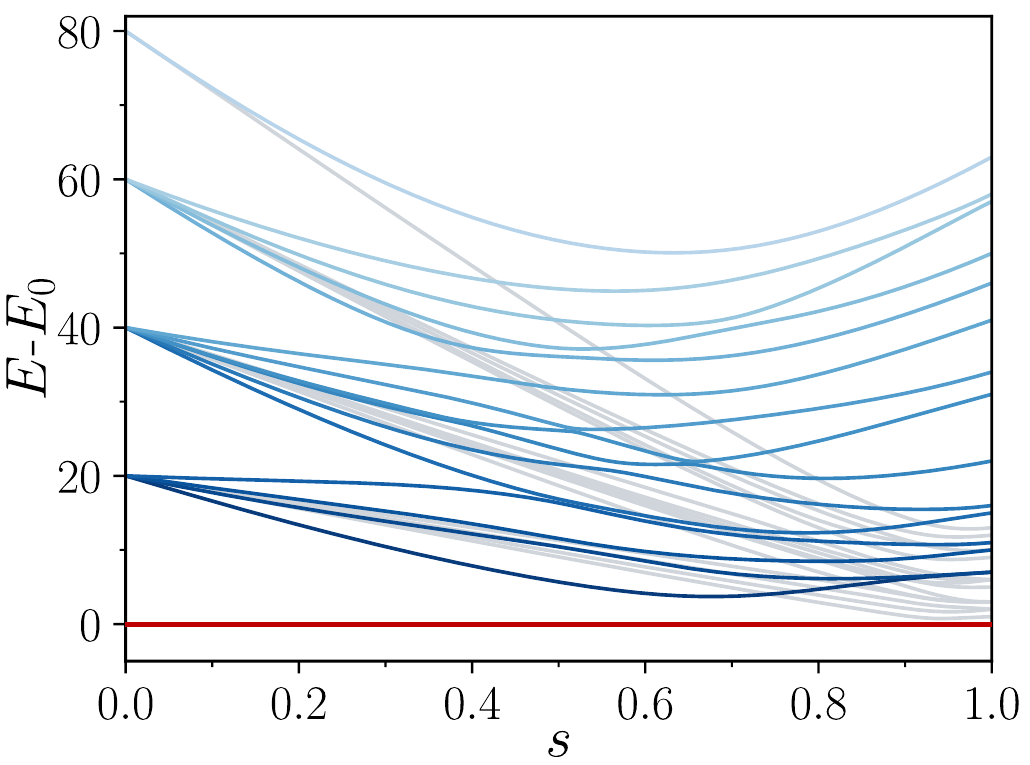}
  \caption{A comparison of the energy spectrum of the total Hamiltonian $H_p$ and $\hat H^\prime_p$ for factorizing 2479. The separation of the energy eigenvalues from the lowest energy level $E_0$ is plotted as a function of $s$, where the grey curves represent the spectrum of $H_p$ and the blue curves represent $\hat H^\prime_p$. The energy difference between the ground state and the first excited state of $\hat H^\prime_p$ is enlarged.
}
\label{2479_spectrum}
\end{figure} 

\begin{figure} 
  \centering
  \includegraphics[width=1\columnwidth]{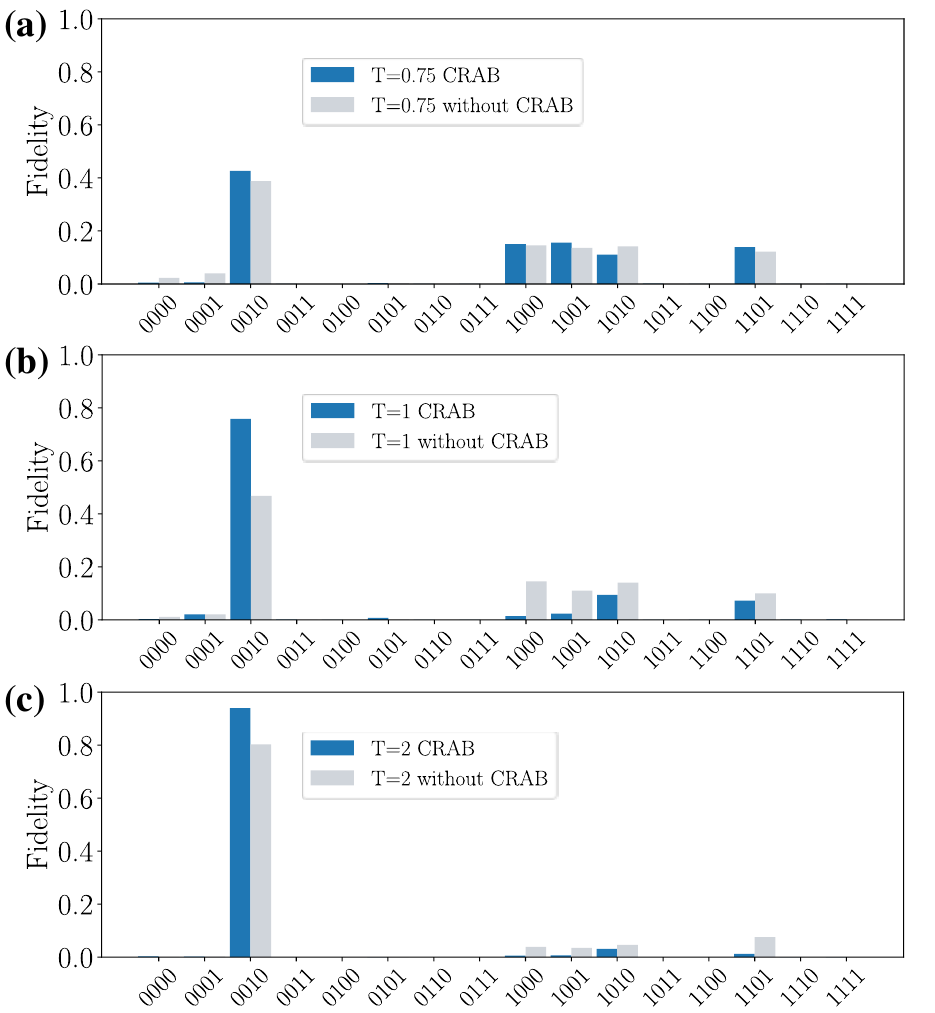}
  \caption{Fidelity of obtaining the ground state under different total evolution time with the CRAB optimization (blue) and initial guess $s_0(t)$ (gray) for factoring 2479. In all cases the ground state $\left|0010\right\rangle$ exhibits the highest fidelity, which is further enhanced by the application of CRAB optimization.}
  \label{Fig4}
\end{figure} 


To enhance the efficiency of the algorithm, whose runtime $T$ is limited by the minimum energy gap, 
we aim to increase the energy difference between the ground state and the first excited state without altering the final state~\cite{zhuang14increase,Omran2019Generation}.
This is accomplished by modifying the coefficients of the bitwise Hamiltonian, resulting in a refined problem Hamiltonian:
\begin{eqnarray}
\hat H^\prime_p &=& (\hat a_3 \hat b_1-\hat b_1)^2 +(\hat a_3 \hat b_2-\hat b_1)^2 \nonumber\\
 && + (\hat a_3+\hat b_2+\hat c_{7,8}-\hat I)^2 + 10(\hat b_1-\hat b_2-2\hat c_{7,8}+\hat I)^2\nonumber\\
 && + 5(\hat a_3-2\hat b_1 \hat b_2-\hat b_1+\hat b_2-\hat I)^2, \label{2479H_p2}
\end{eqnarray}
This adjustment increases the energy gap, as shown in Fig.~\ref{2479_spectrum}. 
Expanding Eq.~\eqref{2479H_p2}, we obtain:
\begin{eqnarray}
\hat H^\prime_p&=&4.75\sigma_z^{(1)}-10.5\sigma_z^{(2)}-5.5 \sigma_z^{(3)}-0.5\sigma_z^{(4)}\nonumber\\
&&-5.5\sigma_z^{(1)}\sigma_z^{(2)} +0.5\sigma_z^{(1)}\sigma_z^{(3)}+0.5\sigma_z^{(1)}\sigma_z^{(4)}\nonumber\\
&&-0.25\sigma_z^{(2)}\sigma_z^{(3)}-10\sigma_z^{(2)}\sigma_z^{(4)}+10.5\sigma_z^{(3)}\sigma_z^{(4)} \nonumber\\
&&+2.75\sigma_z^{(1)}\sigma_z^{(2)}\sigma_z^{(3)}+29.25\hat I, \nonumber
\end{eqnarray}
where $\sigma_z^{(i)}\sigma_z^{(j)}$ and $\sigma_z^{(i)}\sigma_z^{(j)}\sigma_z^{(k)}$ represent interactions involving two and three qubits, respectively.

Regarding the feasibility of the problem Hamiltonian, 
the two-body interaction, such as $\sigma_z^{(i)}\sigma_z^{(j)}$, 
can be realized using the Molmer-Sorensen two-body Ising interaction~\cite{molmer1999multiparticle}. 
For many-body interactions, such as $\sigma_z^{(i)}\sigma_z^{(j)}\sigma_z^{(k)}$, Katz et al.~\cite{katz2022nbody, shapira2023robust} proposed engineering $N$-body entangling interactions between trapped atomic ion qubits. 
This can be achieved through qubit state-dependent squeezing and displacement forces on the collective atomic motion, which generate full $N$-body interactions.
The theory behind two- and three-body interactions has been successfully implemented in numerous experimental and simulation studies~\cite{katz2023demonstration, katz2023programmable, fang2023realization, chu2023photon, farrell2023preparation, kinos2023optical}, providing strong evidence of their feasibility.

The initial Hamiltonian is given by $\hat H_0 = - g(\hat \sigma_x^{(1)} + \hat \sigma_x^{(2)} + \hat \sigma_x^{(3)}+\hat \sigma_x^{(4)})$. 
We apply the CRAB optimization method to obatin the time-dependent function $f(t)$ for different evolution times $T=0.75, 1.0, 2.0$, respectively. 
Figure~\ref{Fig4} compares the CRAB optimization results with the case of initial guess $s_0$, the probability distribution of the final state across different bases shows significant improvement for the problem solution. 
As the total evolution time increases, the distribution of the state $\left|0010\right\rangle$ becomes more concentrated, indicating that the factorization result corresponds to the state $\left|0010\right\rangle$, which implies $a_3=0$, $b_1=0$, $b_2=1$, $c_{7,8}=0$.  
Calculations of other binary variables are provided in Appendix~\ref{app:A},
which are $$(a_5,a_4,a_3,a_2,a_1)=(0,0,0,0,1),$$ and $$(b_4,b_3,b_2,b_1)=(0,0,1,0).$$ 
These correspond to the factors $$x=2^0+\sum_{k=1}^5 2^k a_k+2^6=67$$ 
and $$y=2^0+\sum_{k=1}^4 2^k b_k+2^5=37.$$
Hence, we conclude that $2479$ factors as $67\times37$.

\subsection{Minimum time required for CRAB optimization}
\begin{figure*}
  \centering
  \includegraphics[width=1.5\columnwidth]{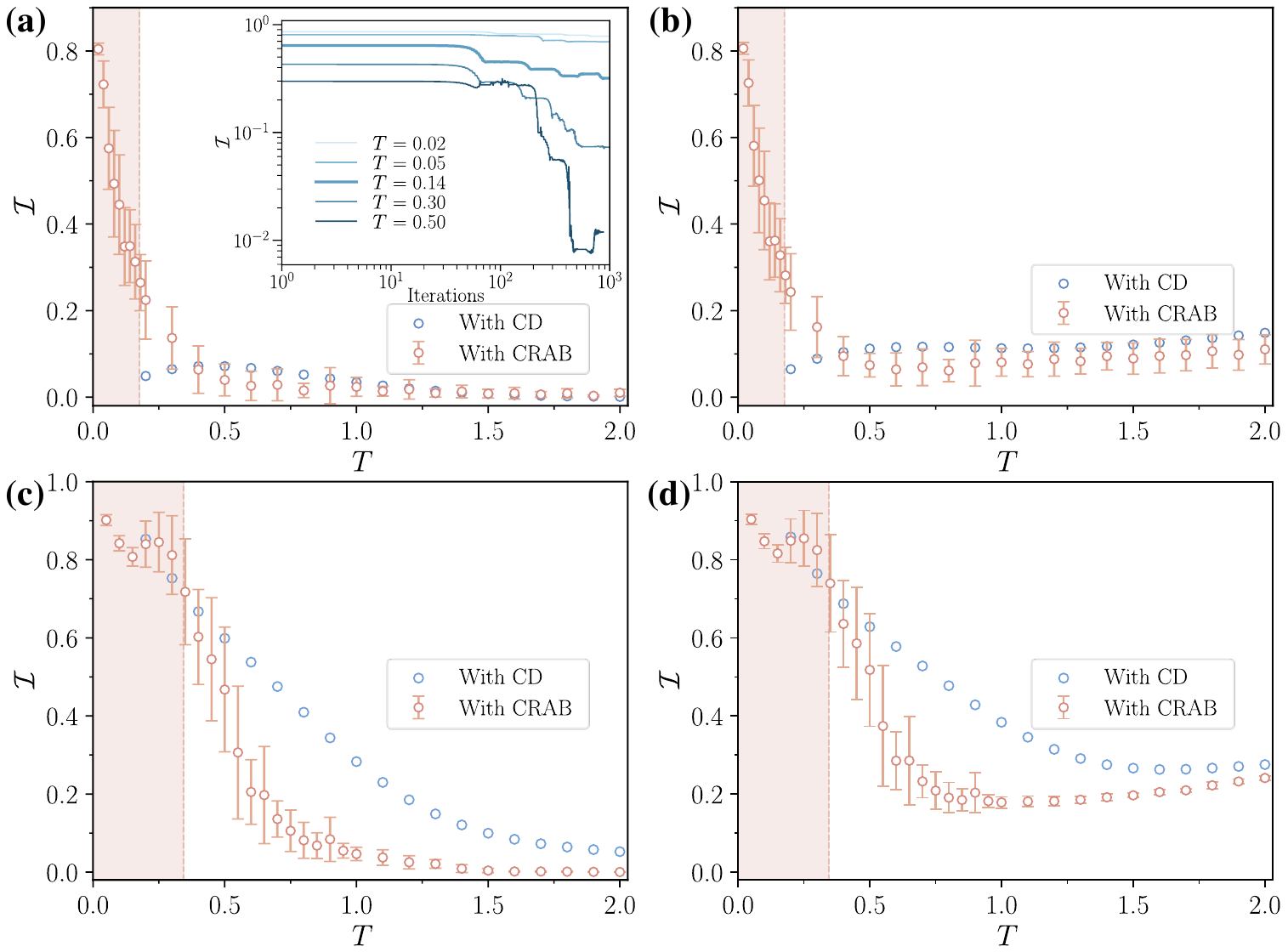}
  \caption{Infidelity $\mathcal{I}$ as a function of total evolution time $T$ for the factorization of 21 and 91. 
  Red points show averages over 10 CRAB runs with different random sets $\{r_k\}$, and error bars indicate standard deviations.
  Blue points represent results from the local CD method.
  The estimated quantum speed limit $T_{\rm QSL}$, obtained from the energy spectrum, 
  is marked by a vertical dashed line, with the region $T < T_{\rm QSL}$ shaded in light red. 
  (a) Factorization of 21 without noise. Inset: infidelity $\mathcal{I}$ versus optimization iterations for different values of $T$ (from 0.02 to 0.5) in a log-log scale. (b) Factorization of 21 with dephasing noise ($\gamma = 0.04$). 
  (c) Factorization of 91 without noise. 
  (d) Factorization of 91 with dephasing noise ($\gamma = 0.04$).}
\label{21-91CRABvsCD1}
\end{figure*}

In Fig.~\ref{21-91CRABvsCD1}(a), we present the results of CRAB optimization performed for factoring 21,
evaluated over a range of total evolution times $T$. 
Each (red) data point shows the average over multiple random initializations, 
with error bars indicating the standard deviation.
For comparison, we also show the results obtained using the local CD method introduced in Ref.~\cite{hegade2021digitized} (blue circles).

According to Ref.~\cite{hegade2021digitized}, the approximate local CD driving term can be written as:
\begin{equation*}
\hat{H}_{\rm CD}=\dot{s}\tilde{A}_{\lambda} = \dot{s}\sum_i \alpha_i(t) \hat{\sigma}_y^{(i)},
\end{equation*}
where $\alpha_i = {[{h}_z^{(i)} \dot{h}_x^{(i)} - {h}_x^{(i)} \dot{h}_z^{(i)}]}/{[2\dot{s} R_i(t)]}$
and $R_i(t) =  {h_z^{(i)}}^2 + {h_x^{(i)}}^2 + 2\sum_j (J_{ij})^2 + 3\sum_{j<k}(K_{ijk})^2 + 4\sum_{j<k<l} (L_{ijkl})^2$.
The time-dependent parameters are defined as: $h_x^{(i)} = (1 - s)\tilde{h}_x^{(i)}$, 
$h_z^{(j)} =s\tilde{h}_z^{(j)}$, 
$J_{ij} = s\tilde{J}_{ij}$, $K_{ijk} = s\tilde{K}_{ijk}$, 
$L_{ijkl} = s\tilde{L}_{ijkl}$,
arising from the decomposition of the initial Hamiltonian $\hat{H}_0= \sum_i \tilde{h}_x^{(i)} \hat{\sigma}_x^{(i)}$
and the problem Hamiltonian
$
\hat{H}_p =  \sum_{i} \tilde{h}_z^{(i)} \hat{\sigma}_z^{(i)} 
            + \sum_{i<j} \tilde{J}_{ij} \hat{\sigma}_z^{(i)} \hat{\sigma}_z^{(j)} 
            + \sum_{i<j<k} \tilde{K}_{ijk} \hat{\sigma}_z^{(i)} \hat{\sigma}_z^{(j)} \hat{\sigma}_z^{(k)}  + \sum_{i<j<k<l} \tilde{L}_{ijkl} \hat{\sigma}_z^{(i)} \hat{\sigma}_z^{(j)} \hat{\sigma}_z^{(k)} \hat{\sigma}_z^{(l)}.
$

\begin{figure*}[htb]
  \centering
  \includegraphics[width=1.5\columnwidth]{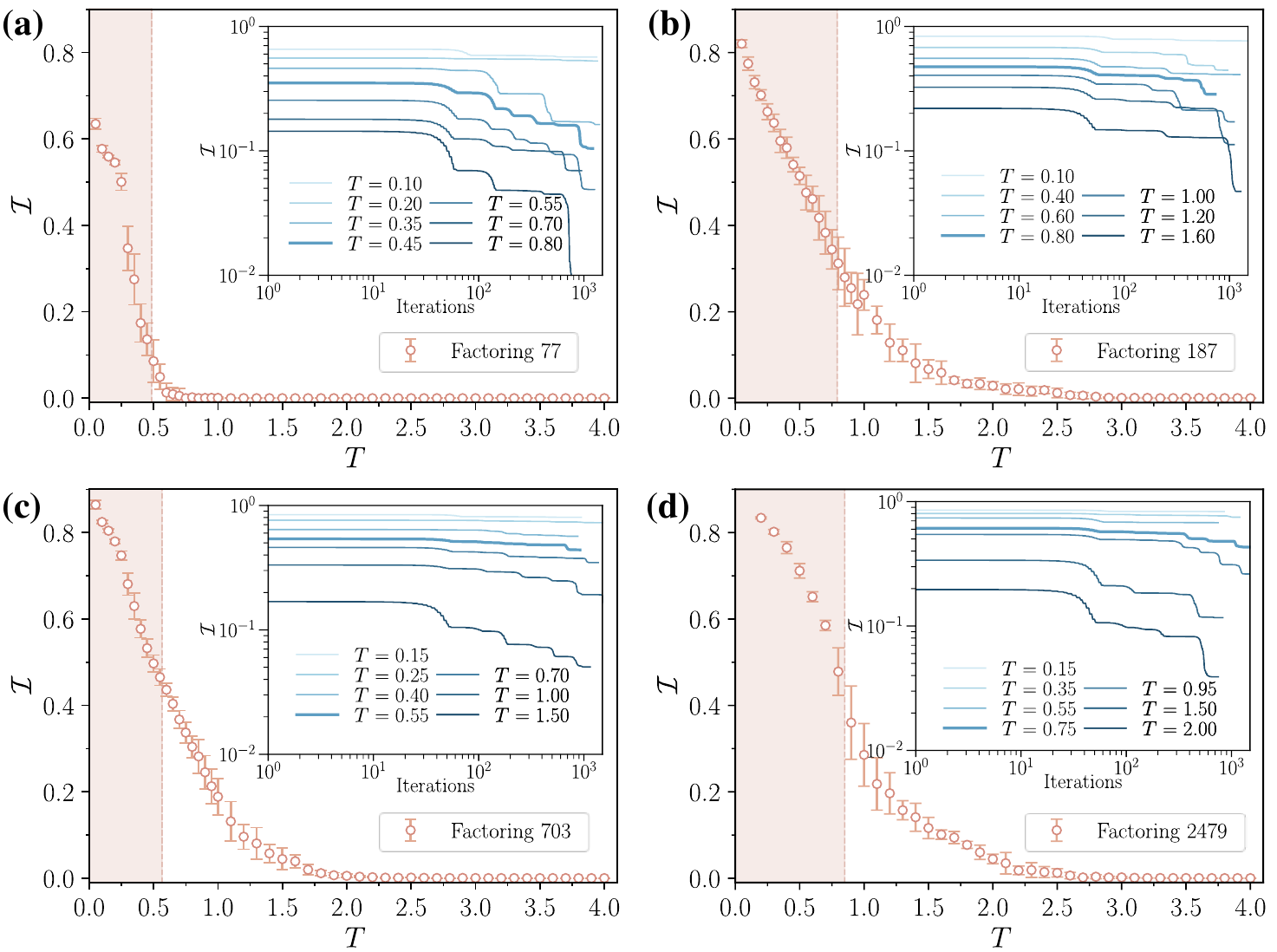}
  \caption{Infidelity $\mathcal{I}$ as a function of total evolution time $T$ for the factorization of (a) 77, (b) 187, (c) 703, and (d) 2479 using CRAB optimization.
  The estimated quantum speed limit $T_{\rm QSL}$, 
  obtained from the energy spectrum, is indicated by vertical dashed lines, 
  with the region $T < T_{\rm QSL}$ highlighted in light red shading. 
  Insets: infidelity $\mathcal{I}$ as a function of optimization iterations for different values of $T$, shown on a log-log scale, the bold curve in each inset corresponds to the threshold evolution time $T_c$. 
  }
\label{77-187-703-2479}
\end{figure*}

In our simulations, we observe a characteristic time scale beyond which the performance of CRAB optimization improves significantly. To interpret this threshold theoretically, we compute the QSL, which provides a lower bound on the time required for a quantum system to evolve between orthogonal states. Using a two-level approximation, where the dynamics are assumed to be dominated by the ground and first excited states, we estimate the QSL as $T_{\rm QSL}=\pi/\Delta_{\rm min}$~\cite{caneva2011speeding}.
For the factorization of the number 21, the minimal spectral gap $\Delta_{\rm min}\approx 17.86$, as shown in Fig.~\ref{fig:crab21}(a), 
yielding $T_{\rm QSL}\approx 0.176$. 
When $T\lesssim0.176$ (highlighted by the shaded region), CRAB performs worse than the local CD method.
As $T$ increases beyond this point, the infidelity of CRAB begins to drop rapidly.
At a slightly later time, it eventually falls below that of the CD method, demonstrating superior performance in the longer-time regime.
Additional insight comes from the inset, which shows the convergence behavior over optimization iterations on a log-log scale. 
There, the transition from stagnation to rapid infidelity decay occurs around $T_c\approx 0.14$,  providing a complementary, empirically defined time-to-solution.
Together, these two time scales, $T_{\rm QSL}\approx 0.176$ from spectral analysis and 
$T_c\approx 0.14$ from convergence behavior, 
consistently identify the emergence of efficient dynamics. 
Despite the underlying two-level approximation, the QSL provides a meaningful and physically grounded estimate that is well supported by our numerical observations.

To further assess the practical relevance of these results, 
we evaluate the robustness of both methods under decoherence. 
In particular, we consider dephasing noise and simulate the system dynamics using the Lindblad master equation:
\begin{equation}
\dot\rho = -i[\hat{H}, \rho]+\gamma\sum_k\mathcal{D}[\hat\sigma_z^k]\rho, \label{eq:mastereq}
\end{equation}
where $\mathcal{D}[\hat{o}]\rho=\hat{o}\rho\hat{o}^\dagger-\frac{1}{2}(\hat{o}^\dagger\hat{o}\rho+\rho\hat{o}^\dagger\hat{o})$ is the Lindblad dissipator.
The noise channel corresponds to pure dephasing on each qubit, with rate $\gamma$. 
As shown in Fig.~\ref{21-91CRABvsCD1}(b), CRAB optimization maintains its performance advantage over the local CD method even in the presence of dephasing, with consistently lower infidelity for $T\gtrsim 0.4$ at a dephasing rate of $\gamma=0.04$.

Figures~\ref{21-91CRABvsCD1}(c) and (d) show the results for the factorization of 91 without and with dephasing noise, respectively. 
In both cases, once the evolution time exceeds the threshold (approximately $T\gtrsim 0.4$), the infidelity of CRAB drops rapidly and subsequently falls below that of the local CD method, consistent with the behavior observed in the 21 case.

To further verify the generality of the observed threshold behavior and its connection to the QSL,
we extend our analysis to additional instances of integer factorization, including 77, 187, 703, and 2479. 
For each case, we analyze both the infidelity as a function of total evolution time and the convergence of CRAB optimization over iterations. 
As shown in Fig.~\ref{77-187-703-2479}, all cases consistently exhibit a transition from inefficient to efficient optimization once the evolution time exceeds a problem-specific threshold $T_c$.
This threshold correlates well with the theoretically estimated QSL $T_{\rm QSL}$,
indicated by vertical dashed lines in the plots. These observations further support the QSL as a meaningful reference timescale for predicting the onset of high-fidelity control across different problem sizes.
For the case of 2479 [Fig.~\ref{77-187-703-2479}(d)], using the Hamiltonian $\hat{H}_p^\prime$
defined in Eq.~\eqref{2479H_p2}, we estimate the QSL as $T_{\rm QSL}\sim 0.85$
and extract a time-to-solution of approximately $T_c\sim 0.75$ from the convergence curves. 
CRAB optimization becomes effective once $T\leq T_c$, consistent with our observations for smaller numbers.  
We note that, unlike the CD-based method in Ref.~\cite{hegade2021digitized}, which saturates at a fidelity of around 0.4 even considering higher-order terms in the CD Hamiltonian, our CRAB optimization achieves significantly higher fidelity beyond the threshold $T_c$.
For 77, 187, and 703 [Figs.~\ref{77-187-703-2479}(a-c)], constructed via binary multiplication table preprocessing using 2, 3, and 4 qubits respectively, details of the Hamiltonian construction and energy spectra are provided in the Appendix. These results further demonstrate that the characteristic time separating inefficient and efficient regimes correlates closely with the QSL in each case.

\section{Conclusions and discussions}\label{sec:conclusion}
In this work, we applied CRAB optimization to enhance adiabatic quantum algorithms for integer factorization. By reformulating the problem as a ground-state search of a suitably constructed Hamiltonian, we demonstrated that CRAB can significantly improve the fidelity of the final state within feasible evolution times. Our results for factoring 21 and 91 show that, even in the presence of decoherence, CRAB outperforms local CD methods when the evolution time exceeds a problem-specific threshold closely related to the QSL. For larger instances such as 2479, our approach achieves higher fidelity than previously reported higher-order CD methods.

These findings highlight the practical value of CRAB-based control strategies for near-term quantum devices. 
The observed correspondence between the QSL and the empirical time-to-solution provides a physically meaningful benchmark for optimization. While our results demonstrate the effectiveness of CRAB, further research is needed to explore its integration with advanced techniques such as COLD-CRAB~\cite{cepaite2024counterdiabatic} and Floquet-CD~\cite{boyers2019floquet}, and to mitigate decoherence in realistic settings. Overall, this study represents a step forward in optimizing adiabatic quantum algorithms through robust and scalable quantum control.

\section*{Acknowledgments}
This work is supported by the Innovation Program for Quantum Science and Technology (Grant No.\;2021ZD0302100), 
the National Natural Science Foundation of China (Grant No. 12304543),
and the Undergraduate Innovation Training Program (No.\;202210287082Y). 
M.X. is also supported by the Jiangsu Province \& Longcheng Youth Science and Technology Talent Support Project.

{\it Data Availability}: The data that support the findings of this study are openly available in Zenodo at \cite{yang_2025_15290975}.

\bibliography{mybib1}


\appendix \label{app:fulu}

\section{quantum factorization of 2479 using the binary multiplication table}
\label{app:A}

The binary multiplication table for factoring 2479 is given in Table~\ref{table:multi_2479},
and the required qubit lengths are $n_a=7$ and $n_b=6$, respectively.
The initial and final qubits are both set to 1. 
The classical preprocessing process for determining variables that are not included in the final set of equations is as follows:

First, consider the equation summing the first column from the left in Table~\ref{table:multi_2479}:
\begin{equation}
1+c_{9,11}+c_{10,11}=1,
\end{equation}
from which we obtain $c_{9,11}=c_{10,11}=0$. Proceeding to the second column from the left, we have
\begin{equation}
b_{4}+a_{5}+c_{9,10}+c_{8,10}-2c_{10,11}=0,\nonumber
\end{equation}
leading to $$b_{4}=a_{5}=c_{9,10}=c_{8,10}=0.$$ 
Similarly, for the third column and the fourth column, the equations are:
\begin{eqnarray}
b_{3}+b_{4}a_{5}+a_{4}+c_{8,9}+c_{7,9}-2c_{9,10}-4c_{9,11}&=&0,\nonumber\\
b_{2}+b_{3}a_{5}+b_{4}a_{4}+a_{3}+c_{7,8}+c_{6,8}-2c_{8,9}-4c_{8,10}&=&1,\nonumber 
\end{eqnarray}
from which we deduce that $$b_{3}=a_{4}=c_{8,9}=c_{7,9}=0,$$ and $$b_{2}+a_{3}+c_{7,8}+c_{6,8}=1.$$

Examining the remaining columns in Table~\ref{table:multi_2479}, we derive other equations, from which we can reduce the problem to a set of equations involving only 4 independent binary variables as follows:
\begin{eqnarray}
a_3b_1-b_1 &=& 0,\\
a_3b_2-b_1 &=& 0,\\
a_3+b_2+c_{7,8}-1&=&0,\\
b_1-b_2-2c_{7,8}+1&=&0,\\
a_3-2b_1b_2-b_1+b_2-1&=&0.
\end{eqnarray}

By squaring and summing all the equations, we get the cost function as
\begin{equation}
  \begin{aligned}
  f(a_3,b_1,b_2,c_{7,8})=(a_3 b_1-b_1)^2 +(a_3 b_2-b_1)^2\\
  +(a_3+b_2+c_{7,8}-1)^2 +(b_1-b_2-2c_{7,8}+1)^2\\
  +(a_3-2b_1b_2-b_1+b_2-1)^2.
  \end{aligned}
\end{equation}
The minimum of this cost function $f_{min}(x, y, c) = 0$. By mapping the binary variables to the qubit operator, we obtained the
final Hamiltonian given in Eq.~(\ref{2479H_p1}) in the main manuscript.

\section{quantum factorization of 703 using the binary multiplication table}
\label{app:703}

The binary multiplication table for factoring 703 is given in Table~\ref{table:multi_703},
and the required qubit lengths are $n_a=6$ and $n_b=5$, respectively.
The initial and final qubits are both set to 1. 
Using the classical preprocessing process, we can finally achieve a set of equations involving only 4 independent binary variables as follows:
\begin{eqnarray}
a_1+a_2+a_3-2a_1a_2-1 &=& 0,\\
a_3-a_1a_3 &=& 0,\\
a_3-a_2a_3+a_1-2c_{5,6}&=&0,\\
1-a_1-a_2+2a_3+c_{5,6}&=&0.
\end{eqnarray}

By squaring and summing all the equations, we get the cost function as
\begin{equation}
  \begin{aligned}
  f(a_1,a_2,a_3,c_{5,6})=(a_1+a_2+a_3-2a_1a_2-1)^2 \\
  +(a_3-a_1a_3)^2+(a_3-a_2a_3+a_1-2c_{5,6})^2 \\+(1-a_1-a_2+2a_3+c_{5,6})^2
  \end{aligned}
\end{equation}

The minimum of this cost function $f_{min}(x, y, c) = 0$. By mapping the binary variables to the qubit operator, we can obtain the final Hamiltonian. In the same way as in Eq.~(\ref{2479H_p2}), we can also adjust the coefficients to increase the energy gap for better Hamiltonian.For example, we can try the cost function as:
\begin{equation}
  \begin{aligned}
  f(a_1,a_2,a_3,c_{5,6})=(a_1+a_2+a_3-2a_1a_2-1)^2 \\
  +(a_3-a_1a_3)^2+10(a_3-a_2a_3+a_1-2c_{5,6})^2 \\+5(1-a_1-a_2+2a_3+c_{5,6})^2
  \end{aligned}
\end{equation}
so the Hamiltonian $\hat{H}_p$ can be:
\begin{equation}
  \begin{aligned}
  \hat{H}_p=(\hat{a}_1+\hat{a}_2+\hat{a}_3-2\hat{a}_1\hat{a}_2-1)^2 +(\hat{a}_3-\hat{a}_1\hat{a}_3)^2\\+10(\hat{a}_3-\hat{a}_2\hat{a}_3+\hat{a}_1-2\hat{c}_{5,6})^2 \\+5(1-\hat{a}_1-\hat{a}_2+2\hat{a}_3+\hat{c}_{5,6})^2
  \end{aligned}\label{eq:703}
\end{equation}

\section{quantum factorization of 187 using the binary multiplication table}
\label{app:187}
The binary multiplication table for factoring 187 is given in Table~\ref{table:multi_187},
and the required qubit lengths are $n_a=5$ and $n_b=4$, respectively.
The initial and final qubits are both set to 1. 
Using the classical preprocessing process, we can finally achieve a set of equations involving only 3 independent binary variables as follows:
\begin{eqnarray}
a_1+b_1-1 &=& 0,\\
a_1-2c_{4,5}&=&0,\\
b_1+c_{4,5}-1 &=& 0.
\end{eqnarray}

By squaring and summing all the equations, we get the cost function as
\begin{equation}
  \begin{aligned}
  f(a_1,b_1,c_{4,5})=(a_1+b_1-1)^2+(a_1-2c_{4,5})^2 
  \\+(b_1+c_{4,5}-1)^2
  \end{aligned}
\end{equation}
The minimum of this cost function $f_{min}(x, y, c) = 0$. By mapping the binary variables to the qubit operator, we can obtain the final Hamiltonian. In the same way as in Eq.~(\ref{2479H_p2}), we can also adjust the coefficients to increase the energy gap for better Hamiltonian.For example, we can try the cost function as:
\begin{equation}
  \begin{aligned}
  f(a_1,b_1,c_{4,5})=(a_1+b_1-1)^2+5(a_1-2c_{4,5})^2 \\
  +10(b_1+c_{4,5}-1)^2
  \end{aligned}
\end{equation}
so the Hamiltonian $\hat{H}_p$ can be:
\begin{equation}
  \begin{aligned}
  \hat{H}_p=(\hat{a}_1+\hat{b}_1-1)^2+5(\hat{a}_1-2\hat{c}_{4,5})^2 \\
  +10(\hat{b}_1+\hat{c}_{4,5}-1)^2
  \end{aligned}\label{eq:187}
\end{equation}
\section{quantum factorization of 77 using the binary multiplication table}
\label{app:77}
The binary multiplication table for factoring 77 is given in Table~\ref{table:multi_77},
and the required qubit lengths are $n_a=4$ and $n_b=3$, respectively.
The initial and final qubits are both set to 1. 
Using the classical preprocessing process, we can finally achieve a set of equations involving only 2 independent binary variables as follows:
\begin{eqnarray}
a_1+2a_2-1 &=& 0,\\
a_1+a_2-1&=&0.
\end{eqnarray}

By squaring and summing all the equations, we get the cost function as
\begin{equation}
  \begin{aligned}
  f(a_1,a_2)=(a_1+2a_2-1)^2+(a_1+a_2-1)^2
  \end{aligned}
\end{equation}
The minimum of this cost function $f_{min}(x, y, c) = 0$. By mapping the binary variables to the qubit operator, we can obtain the final Hamiltonian. In the same way as in Eq.~(\ref{2479H_p2}), we can also adjust the coefficients to increase the energy gap for better Hamiltonian. For example, we can try the cost function as:
\begin{equation}
  \begin{aligned}
  f(a_1,a_2)=10(a_1+2a_2-1)^2 +5(a_1+a_2-1)^2
  \end{aligned}
\end{equation}
so the Hamiltonian $\hat{H}_p$ can be:
\begin{equation}
  \begin{aligned}
  \hat{H}_p=10(\hat{a}_1+2\hat{a}_2-1)^2 +5(\hat{a}_1+\hat{a}_2-1)^2
  \end{aligned}\label{eq:77}
\end{equation}

\begin{table*}
  \small
  \centering
  \caption{Multiplication table for 2479 in binary}
  \begin{ruledtabular}
  \begin{tabular}{ccccccccccccc}
 & $2^{11}$& $2^{10}$ & $2^9$ & $2^8$& $2^7$& $2^6$ & $2^5$ & $2^4$ & $2^3$& $2^2$ & $2^1$ & $2^0$ \\
$a$&&&&&&1&$a_5$&$a_4$&$a_3$&$a_2$&$a_1$& 1 \\
$b$&&&&&&&1&$b_4$& $b_3$& $b_2$ & $b_1$ & 1 \\
  \colrule
  &&&&&&1&$a_5$& $a_4$&$a_3$&$a_2$&$a_1$& 1 \\
  &&&&&$b_1$&$b_1a_5$&$b_1a_4$&$b_1a_3$& $b_1a_2$& $b_1a_1$ & $b_1$ &\\
  &&&&$b_2$&$b_2a_5$&$b_2a_4$&$b_2a_3$& $b_2a_2$& $b_2a_1$ & $b_2$&&\\
  &&&$b_3$&$b_3a_5$&$b_3a_4$&$b_3a_3$& $b_3a_2$& $b_3a_1$ & $b_3$&&&\\
  &&$b_4$&$b_4a_5$&$b_4a_4$&$b_4a_3$& $b_4a_2$& $b_4a_1$ & $b_4$&&&&\\
    &1&$a_5$& $a_4$&$a_3$&$a_2$&$a_1$& 1&&&&&\\
 carriers&$c_{10,11}$&$c_{9,10}$&$c_{8,9}$&$c_{7,8}$&$c_{6,7}$&$c_{5,6}$&$c_{4,5}$&$c_{3,4}$&$c_{2,3}$& $c_{1,2}$& &  \\
  &$c_{9,11}$&$c_{8,10}$&$c_{7,9}$&$c_{6,8}$&$c_{5,7}$&$c_{4,6}$&$c_{3,5}$&$c_{2,4}$&&&& \\
   \colrule
   $ab=2479$& 1 & 0 & 0 & 1 & 1 & 0 & 1 & 0 & 1 & 1 & 1 & 1
  \end{tabular}
  \end{ruledtabular}\label{table:multi_2479}
\end{table*}

\begin{table*}
  \small
  \centering
  \caption{Multiplication table for 77 in binary}
  \begin{ruledtabular}
  \begin{tabular}{cccccccc}
 & $2^6$ & $2^5$ & $2^4$ & $2^3$& $2^2$ & $2^1$ & $2^0$ \\
$a$&&&&1&$a_2$&$a_1$& 1 \\
$b$&&&&&1& $b_1$ & 1 \\
  \colrule
  &&&&1&$a_2$&$a_1$& 1 \\
  &&&$b_1$& $b_1a_2$& $b_1a_1$ & $b_1$ &\\
  &&1&$a_2$&$a_1$& 1&&\\
 carriers&$c_{5,6}$&$c_{4,5}$&$c_{3,4}$&$c_{2,3}$& $c_{1,2}$& &  \\
  &$c_{4,6}$&$c_{3,5}$&$c_{2,4}$&&&& \\
   \colrule 
   $ab=77$& 1 & 0 & 0 & 1 & 1 & 0 & 1
  \end{tabular}
  \end{ruledtabular}\label{table:multi_77}
\end{table*}

\begin{table*}
  \small
  \centering
  \caption{Multiplication table for 187 in binary}
  \begin{ruledtabular}
  \begin{tabular}{ccccccccc}
 & $2^7$& $2^6$ & $2^5$ & $2^4$ & $2^3$& $2^2$ & $2^1$ & $2^0$ \\
$a$&&&&1&$a_3$&$a_2$&$a_1$& 1 \\
$b$&&&&&1& $b_2$ & $b_1$ & 1 \\
  \colrule
  &&&&1& $a_3$&$a_2$&$a_1$& 1 \\
  &&&$b_1$&$b_1a_3$& $b_1a_2$& $b_1a_1$ & $b_1$ &\\
  &&$b_2$&$b_2a_3$& $b_2a_2$& $b_2a_1$ & $b_2$&&\\
  &1&$a_3$&$a_2$&$a_1$& 1&&\\
 carriers&$c_{6,7}$&$c_{5,6}$&$c_{4,5}$&$c_{3,4}$&$c_{2,3}$& $c_{1,2}$& &  \\
  &$c_{5,7}$&$c_{4,6}$&$c_{3,5}$&$c_{2,4}$&&&& \\
   \colrule 
   $ab=187$& 1 & 0 & 1 & 1 & 1 & 0 & 1 & 1 
  \end{tabular}
  \end{ruledtabular}\label{table:multi_187}
\end{table*}

\begin{table*}
  \small
  \centering
  \caption{Multiplication table for 703 in binary}
  \begin{ruledtabular}
  \begin{tabular}{ccccccccccc}
 & $2^9$ & $2^8$& $2^7$& $2^6$ & $2^5$ & $2^4$ & $2^3$& $2^2$ & $2^1$ & $2^0$ \\
$a$&&&&&1&$a_4$&$a_3$&$a_2$&$a_1$& 1 \\
$b$&&&&&&1& $b_3$& $b_2$ & $b_1$ & 1 \\
  \colrule
  &&&&&1& $a_4$&$a_3$&$a_2$&$a_1$& 1 \\
  &&&&$b_1$&$b_1a_4$&$b_1a_3$& $b_1a_2$& $b_1a_1$ & $b_1$ &\\
  &&&$b_2$&$b_2a_4$&$b_2a_3$& $b_2a_2$& $b_2a_1$ & $b_2$&&\\
  &&$b_3$&$b_3a_4$&$b_3a_3$& $b_3a_2$& $b_3a_1$ & $b_3$&&&\\
  &1& $a_4$&$a_3$&$a_2$&$a_1$& 1&&&&\\
 carriers&$c_{8,9}$&$c_{7,8}$&$c_{6,7}$&$c_{5,6}$&$c_{4,5}$&$c_{3,4}$&$c_{2,3}$& $c_{1,2}$& &  \\
  &$c_{7,9}$&$c_{6,8}$&$c_{5,7}$&$c_{4,6}$&$c_{3,5}$&$c_{2,4}$&&&& \\
   \colrule 
   $ab=703$& 1 & 0 & 1 & 0 & 1 & 1 & 1 & 1 & 1 & 1 
  \end{tabular}
  \end{ruledtabular}\label{table:multi_703}
\end{table*}

\begin{figure*}
  \centering
  \includegraphics[width=1.9\columnwidth]{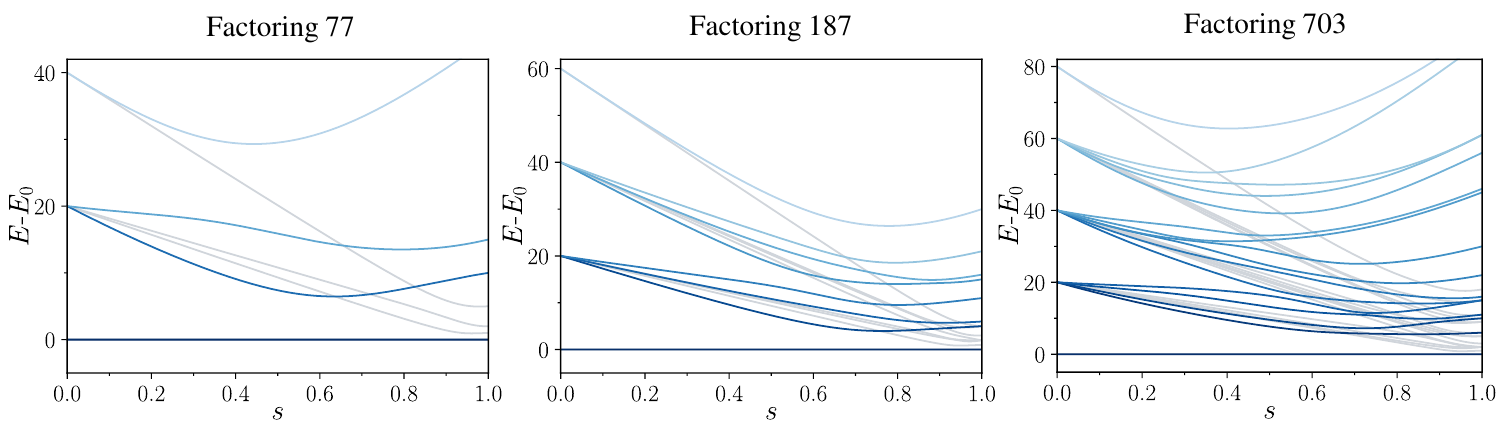}
  \caption{The energy spectrum of Hamiltonian for factoring 77, 187 and 703}
\label{77-187-703-gap}
\end{figure*}

\end{document}